\documentstyle[12pt]{article}
\def \be{\begin{equation}}
\def \ee{\end{equation}}                
\def \ba{\begin{array}{l}}
\def \ea{\end{array}}
\def \bq{\begin{eqnarray}}
\def \eq{\end{eqnarray}}
\def \nn{\nonumber\\}
\def \lb{\label}
\def \ln{{\rm ln}}

\def \fr{\frac}
\def \Tr{{\rm Tr}}
\def \tl{\tilde}
\def \ol{\overline}

\def \la{\langle}
\def \ra{\rangle}
\def \[{\left[}
\def \]{\right]}
\def \({\left(}
\def \){\right)}
\def \2{\frac{1}{2}}
\def \4{\frac{1}{4}}

\def \g{\gamma}

\def \d{\delta}
\def \D{\Delta}
\def \e{\epsilon}
\def \f{\phi}
\def \lm{\lambda}
\def \Lm{\Lambda}
\def \n{\nabla}
\def \p{\varphi}

\def \t{\tau}

\def \T{T_{c}}
\def \I{\int d^{D}x}
\def \Ip{\int_{|p|<1}\frac{d^{D}p}{(2\pi)^{D}}}

\newcommand{\sectio}[1]{\section{#1}\setcounter{equation}{0}}

\begin{document}

\frenchspacing
\setlength{\parskip}{2mm}

\vspace{10mm}

\begin{center}

{\Large \bf Is There a Spin-Glass Phase 
            in the Random Temperature Ising Ferromagnet?}
\vskip .2in
Gilles Tarjus and Victor Dotsenko\footnote{On leave from 
Landau Institute for Theoretical Physics, Moscow}
\vskip .1in
Laboratoire de Physique  Theorique des Liquides, \\
UMR 7600, Universite Paris VI, \\
4 place Jussieu, 75252 Paris Cedex 05,
France\\

\end{center}

\vskip .3in

\begin{abstract}
In this paper we study the phase diagram of the disordered 
Ising ferromagnet. Within the framework of the Gaussian variational
approximation it is shown that in systems with a finite value of the
disorder in dimensions $D=4$ and $D < 4$ the paramagnetic and 
ferromagnetic phases are separated by a spin-glass phase. 
The transition from paramagnetic to spin-glass state is continuous 
(second-order), while the transition between spin-glass and ferromagnetic
states is discontinuous (first-order). It is also shown that within 
the considered approximation there is no replica symmetry breaking 
in the spin-glass phase. The validity of the Gaussian variational 
approximation for the present problem is discussed, and we provide 
a tentative physical interpretation of the results.
\end{abstract}

\newpage

\sectio{Introduction}

Phase transitions in the random temperature Ising ferromagnets have been 
intensively studied theoretically, numerically and experimentally
during the last decades. The theoretical interest has mainly been focused 
on the critical behavior in the vicinity the 
paramagnetic-ferromagnetic phase transition point $\T$ in {\it weakly} 
disordered systems \cite{crit1}.
Renormalization group considerations show that if the temperature is not too close 
to $\T$,  the critical behavior is essentially controlled by the fixed point
of the pure system
(so that disorder produces only irrelevant corrections), while
in the close vicinity of $\T$ the critical behavior turns out to be different
from that of the pure system and is characterized
by a new universal (independent of the disorder strength) fixed point.

On these grounds it is widely believed that the critical behavior of the disordered
system  is {\it universal}, and the strength of the disorder only affects
the size of the critical region near $\T$ (but not the critical 
behavior itself). In other words, the critical behavior of systems
with a finite value of the disorder must be the same as that in the weakly disordered
ones. Most of the numerical simulations (in particular for the two-dimensional
systems) support this idea, see e.g. \cite{numer1}, although some of the numerical
results seem to indicate that the critical behavior can be non-universal and
characterized by critical exponents depending on the
disorder strength \cite{numer2}.

In this paper we consider the problem of the phase transitions in the 
disordered Ising ferromagnet from a somewhat different point of view. 
Instead of studying of the critical behavior, we propose first to address
a simpler point concerning the nature of the phases in such  
systems. We stress that the usual
assumption that the random temperature Ising ferromagnet can only be
in the paramagnetic or in the ferromagnetic state is, to the least, 
questionable, a point first raised by Ma and Rudnick \cite{Ma-R}.
Before studying such details as the critical properties 
 one should first clarify what kind of phases and what kind of transitions 
can exist in such a system.

A general reason for asking such a question comes from the fact that the
saddle-point equations which describe the local minima of the disordered
Hamiltonian in the 
(supposed) paramagnetic region have an exponentially large number of solutions.
Physically this situation is quite clear: due to the spatial fluctuations of
the local transition temperatures one can find a macroscopic number
of "ferromagnetic islands", well separated in space, that spend most of the time 
in state with a
non-zero local magnetization which can be either positive or negative. 
As long as  these islands are rare
(i.e., away from the supposed ferromagnetic
transition temperature), they lead to the existence of an exponential number
of local minima. Moreover, the presence of rare exponentially large islands 
results in 
the existence of non-analytic (Griffith-like) contributions to the
thermodynamic functions \cite{grif}.

An indirect indication that the phase behavior of
such systems could be more complicated than that described by
the renormalization group
has been obtained in the framework of the so-called non-perturbative
renormalization group (RG) approach \cite{rg-rsb}. In this latter,  
the existence of many different local minima 
of the disordered Hamiltonian is taken into account in the 
form of a replica symmetry breaking scheme,
and it was eventually found that the renormalization flow  
leads to the strong coupling regime at the {\it finite} spatial scale, 
and not to the expected fixed points. 
This may indicate that something is basically wrong with the supposed
(trivial) minimum of the renormalized Hamiltonian.

It has been suggested that when lowering the temperature, the localized
ferromagnetic islands become close and strongly interacting, which leads to a
transition to the global ferromagnetic state \cite{percol}.
The solution of the saddle-point equations 
within the Gaussian variational approximation and the replica framework
described in the
next sections shows that this is not the only possibility.
We indeed find that upon lowering the temperature the global state of the system 
can become a spin glass {\it before} the ferromagnetic 
state sets in.
In this spin-glass state  the total magnetization remains zero, and there is an
effective freezing of local (random) spin configurations (which leads to a
non-zero value of the spin-glass Edwards-Anderson order parameter).
Besides, one finds that in this spin-glass state the two-point spin-spin
correlation function is described by a temperature independent finite 
correlation length (it is interesting to note that this length coincides
with that at which the strong coupling regime of the RG approach \cite{rg-rsb}
sets in), 
while the (spin-glass type) four-spin correlation function
becomes critical at the spin-glass phase transition point.
Finally, when further lowering the temperature the global ferromagnetic
state eventually sets in via a {\it first-order} phase transition. 

The existence of an intermediate spin-glass phase in a system where
a priori no frustrations, no competition of interactions occur
is puzzling. This was stressed by Sherrington \cite{DS} in a response to the
perturbative analysis (not within the framework of the replica method)
of Ma and Rudnick \cite{Ma-R} that predicted such a spin-glass phase.
Besides the potential flaws associated with the perturbative
treatments, the problem lies in the fact that it is hard to
imagine a disordered ferromagnrt in a state where the 4-spin
spin-glass susceptibility is larger than the square of the 2-spin
ferromagnetic susceptibility \cite{DS} nor in a state with a zero
total magnetization and a non-zero spin-glass order parameter.
This will be discussed later, but, at the level of "hand-waving 
arguments", one can propose a possible interpretation
for the presence of a spin-glass phase.
 The point is that as one
lowers the temperature (from the paramagnetic phase side) and the 
ferromagnetic islands become close and strongly interacting, 
there need not be the appearance of a unique infinite (percolating) 
ferromagnetic island.
The existence of the spin-glass solution of the saddle-point
equations in the considered disordered Ising ferromagnet requires that a large
number of  effectively independent spanning
ferromagnetic clusters appear in the system. Just below the transition
each of the clusters is characterized by non-zero value of its own
global magnetization (so that within the cluster the spins are effectively
"frozen"), but the sign of these magnetizations remains random from cluster
to cluster. This situation manifests itself as the spin-glass state with
"frozen" spins and no (averaged over clusters) global magnetization.
Finally, when the temperature is further decreased, the effective
interactions among these spanning clusters become strong enough for the system
to eventually make a "jump" (via a first-order transition) 
into the ferromagnetic state.
In other words, the ferromagnetic phase sets in due to a collective locking of
the orientations of the clusters magnetisations
in the same direction.
It is easy to understand that this  transition must be
first-order. Indeed, since at the point of the spin-glass to ferromagnetic 
transition 
the absolute value of the (randomly directed) magnetizations 
of the spanning ferromagnetic clusters in the spin-glass phase is already
finite,  the value of the global ferromagnetic order parameter
resulting from the locking of the various orientations in the same direction
is itself finite.

In the next section we present 
the general formalism in terms of the standard replica approach 
and of the Gaussian variational
approximation \cite{gva},\cite{gva1} as applied to the random temperature model. 
In Section 3 we derive all the solutions of the corresponding saddle-point equations,
solutions that describe the different "ground states" that can exist in the model. 
It is  shown 
that the spin-glass solution discussed above can exist only in dimensions $D \leq 4$. 
Since the conclusions of the present study are, to a large extent, only of
qualitative nature, we focus on the system in 
dimension $D=4$ (the generalization of the results for dimensions
$D = (4 - \e)$ is given in the Appendix C). We obtain the solutions for the
paramagnetic, (replica-symmetric) spin-glass and ferromagnetic states, and we derive
the temperature regions over which these phases are stable as well as the nature
of the phase transitions separating these these phases.
In section 4 the singularity in the spin-glass-type four-spin correlation function
and in the corresponding susceptibility at the spin-glass phase transition
is derived. (In Appendix A we give the formal proof that, in the framework of the
present formalism, no replica symmetry breaking
solutions, either continuous or step-like, can exist in the spin-glass state.)
Finally, in section 5 we discuss the validity of  the Gaussian variational 
method;  we stress, in particular, that the present
approach can only be reliable for finite values of the parameter describing 
the disorder strength. We also suggest a possible scenario for the existence 
of an intermadiate spin-glass phase.

\sectio{General formalism}

In this paper we study the disordered (random temperature)
$D$-dimensional Ising ferromagnet which can be 
described in the continuous by the following Ginsburg-Landau Hamiltonian:

\be
\lb{r1}
H\[\f(x);\d\t(x)\] = \I \[ \2 \(\n\f(x)\)^{2} + \2 \(\t - \d\t(x)\) \f^{2}(x) +
 \4 g \f^{4}(x) \] .
\ee
Here, $\t \equiv (T - \T)/\T \ll 1$ is the reduced temperature, and the quenched 
disorder is described by random spatial fluctuations of the local transition 
temperature $\d\t(x)$ whose probability distribution is taken to be symmetric 
and Gaussian:

\be
\lb{r2}
P[\d\t] = p_{0} \exp \Biggl( -\frac{1}{4u}\I (\d\t(x))^{2} \Biggr) \; ,
\ee
where $u$ is the parameter which describes the strength of the disorder
and $p_{0}$ is an irrelevant normalization constant. 

In terms of the standard replica method, the averaged (over quenched disorder) free
energy is calculated from an the annealed average involving $n$ copies of the
same system:

\be 
\lb{r3}
F = - \ol{\(\ln Z\)} = - \lim_{n\to 0} \fr{1}{n} \ln\[ \ol{Z^{n}} \]
\ee
where $\ol{(...)}$ denotes the averaging over the random function $\d\t(x)$ 
with the probability distribution (\ref{r2}), and 

\be
\lb{r4}
\ol{Z^{n}} \equiv \int {\cal D}\d\t(x) P[\d\t] 
\[\int{\cal D}\f(x)\exp\( -H\[\f(x);\t(x)\] \) \]^{n}
\ee
is the replica partition function. 
Simple Gaussian integration over $\d\t(x)$ in eq.(\ref{r4}) yields

\be
\lb{r5}
 \ol{Z^{n}} = \prod_{a=1}^{n}\[\int {\cal D}\f_{a}(x)\] \exp\( -H^{(n)}[\f_{a}(x)] \)
\ee
where

\be
\lb{r6}
H^{(n)}[\f_{a}(x)] = 
   \I \[ \2 \sum_{a=1}^{n}\(\n\f_{a}\)^{2} + \2 \t \sum_{a=1}^{n} \f_{a}^{2} +
 \4 \sum_{a,b=1}^{n} g_{ab}\f_{a}^{2} \f_{b}^{2} \]
\ee
is the replica Hamiltonian and 

\be
\lb{r7}
g_{ab} = g \d_{ab} - u  \; \; .
\ee

To take into account the possibility of ferromagnetic ordering in the system
we explicitly introduce the ferromagnetic order parameter $m = \ol{\la\f\ra}$
by redefining the fields as follows:

\be
\lb{r8}
\f_{a}(x) = m + \p_{a}(x) \; ,
\ee
where the new fields $\p_{a}(x)$ describe the spatial fluctuations with
zero mean. By substituting (\ref{r8}) into eq.(\ref{r6}) for the replica 
Hamiltonian, one finds

\bq
\lb{r9}
&&H^{(n)}[\p_{a}(x); m] = V n (\2 \t m^{2} + \4 g m^{4} )  +                                  
     \I \[ \2 \sum_{a=1}^{n}\(\n\p_{a}\)^{2} + \2 \t \sum_{a=1}^{n} \p_{a}^{2} +
       \4 \sum_{a,b=1}^{n} g_{ab}\p_{a}^{2} \p_{b}^{2}    +   \right.                             
\nn
\nn
&& \left. +  \2 g m^{2} \sum_{a=1}^{n}\p_{a}^{2} + m^{2} \sum_{a,b=1}^{n} g_{ab}\p_{a}\p_{b}   
    + \t m  \sum_{a=1}^{n}\p_{a} + m \sum_{a,b=1}^{n} g_{ab}\p_{a}\p_{b}^{2}  +
      m^{3} g \sum_{a=1}^{n}\p_{a}       \] \; \; ,
\eq
where $V$ is the volume of the system. Note that the limit $n \to 0$, that 
must formally be taken in the final results, allows us to omit 
all terms of order $n^{2}$ in the above expression 
(and in further calculations). 

The idea of the Gaussian variational approach is to approximate the fluctuations of the
fields $\p_{a}(x)$ in the above eq.(\ref{r9}) by the Gaussian trial Hamiltonian

\be
\lb{r10}
H^{(n)}_{g}[\p_{a}|{\bf{G}}] = \fr{V}{2} 
      \Ip \sum_{a,b=1}^{n} G_{ab}^{-1}(p)\p_{a}(p) \p_{b}(-p) \; \; ,
\ee
where the correlation functions $G_{ab}(p) = \la\p_{a}(p)\p_{b}(-p)\ra$
are considered as variational parameters.

The replica partition function can be represented as follows:

\be
\lb{r11}
\ol{Z^{n}} = \prod_{a=1}^{n}\[\int {\cal D}\p_{a}(x)\] 
\exp\{ -H^{(n)}_{g}[\p_{a}(x)] - 
(H^{(n)}[\p_{a}(x);m] - H^{(n)}_{g}[\p_{a}(x)] ) \} \; \; ,
\ee
and in the first-order cumulant approximation in the difference 
$(H^{(n)} - H^{(n)}_{g})$ one finds

\be 
\lb{r12}
\ol{Z^{n}} \simeq \exp\[-\2 V \Ip \Tr \; \ln ({\bf{G}}^{-1}(p)) -
   \la (H^{(n)} - H^{(n)}_{g})\ra_{g} \] \equiv
   \exp\( - n V f\[m; {\bf{G}}\] \)
\ee
where $\la(...)\ra_{g}$ denotes the averaging with the Gaussian weight, eq.(\ref{r10});
$ f\[m; {\bf{G}} \]$ is the density of free energy that depends on the order
parameter $m$ and on the trial correlation functions $G_{ab}(p)$:

\be
\lb{r13}
 f\[m; {\bf{G}}\] = \fr{1}{2n} \Ip \Tr \; \ln ({\bf{G}}^{-1}(p)) +
               \fr{1}{nV}\la (H^{(n)} - H^{(n)}_{g})\ra_{g} \; \; .
\ee
Since the above free energy density is an upper bound of the exact replica
free energy density, the variational 
parameters $m$ and $G_{ab}(p)$ can be determined by minimization of eq.(\ref{r13}).
One should however keep in mind the oddities related to the limit
$n \to 0$, in particular the fact that the number of parameters can turn
negative for $n < 1$ (see section 3.3)\footnote{
As noted by Mezard and Parisi in their replica field theory for 
random manifolds \cite{gva1}, the Gaussian variational method
becomes exact when the number $N$ of components of the fields 
$\mbox{\boldmath $\f$}_{a}$ goes to infinity (here, $N=1$). To apply this remark to the
present case, one must generalize the non-Gaussian term appearing in the 
replica Hamiltonian, eq.(\ref{r6}), to
$\fr{1}{12} \sum_{ab=1}^{n}(g_{ab}/N) 
[\mbox{\boldmath $\f$}_{a}^{2}\mbox{\boldmath $\f$}_{b}^{2} + 
2 (\mbox{\boldmath $\f$}_{a} \cdot \mbox{\boldmath $\f$}_{b})^{2} ]  $.
Notice that with this latter term the Hamiltonian {\bf does not} correspond
to the replica-space formulation of the random temperature $O(N)$ model.}.
Inserting eqs.(\ref{r9}) and (\ref{r10}) into eq.(\ref{r13}) leads to

\bq
\lb{r14}
&&f\[m;{\bf{G}}\] = -\fr{1}{2n} \Ip \Tr \; \ln ({\bf{G}}(p)) + \2 \t m^{2} + \4 g m^{4} +
\nn
\nn
&& + \fr{1}{2n}\Ip (p^{2} +\t) \sum_{a=1}^{n} \la\p_{a}(p)\p_{a}(-p)\ra_{g} +
   \fr{1}{4n} \sum_{a,b=1}^{n} g_{ab}\la\p_{a}^{2}(x)\p_{b}^{2}(x)\ra_{g} +
\nn
\nn
&& + \fr{1}{2n} g m^{2} \sum_{a=1}^{n}\la\p_{a}^{2}(x)\ra_{g} +
   \fr{1}{n} m^{2} \sum_{a,b=1}^{n} g_{ab}\la\p_{a}(x) \p_{b}(x)\ra_{g} \; \; .
\eq
Above and in what follows we omit irrelevant constant terms.	
For the Gaussian averages of the fluctuating fields one has

\bq
\lb{r15}
 \la\p_{a}(x)\p_{b}(x)\ra_{g} &=& \Ip G_{ab}(p) \equiv \[G_{ab}\]
\\
\nn 
\lb{r16}
 \la\p_{a}^{2}(x)\ra_{g} &=& \Ip G_{aa}(p) \equiv \[G_{aa}\]
\\
\nn
\lb{r17}
 \la\p_{a}^{2}(x)\p_{b}^{2}(x)\ra_{g} &=& \la\p_{a}^{2}(x)\ra_{g}\la\p_{b}^{2}(x)\ra_{g} +
   2 \la\p_{a}(x)\p_{b}(x)\ra_{g}^{2} \equiv
   \[G_{aa}\]\[G_{bb}\] + 2 \[G_{ab}\]^{2}
\eq
where we have introduced the notation 

\be
\lb{r17a}
\Ip A(p) \equiv \[A\]
\ee
for an arbitrary function $A(p)$.
Taking into account that
the diagonal elements of the matrix ${\bf{G}}$ must be independent of the replica index,
$G_{aa} \equiv \tl{G}$ we find the following expression for the free energy density:

\bq
\lb{r18}
f\[m; {\bf{G}}\] &=& -\fr{1}{2n} \Tr \;\[ \ln ({\bf{G}})\] + \2 \t m^{2} + \4 g m^{4} +
\2\[(p^{2} +\t) \tl{G}\]
\nn
\nn
&& + \4 g \[\tl{G}\]^{2} + \2 (g-u) \[\tl{G}\]^{2} + \fr{1}{2n}\sum_{a\not= b}^{n} g_{ab} \[G_{ab}\]^{2} +
\nn
\nn
&& + \2 g m^{2} \[\tl{G}\] + m^{2} (g-u) \[\tl{G}\] + 
   \fr{1}{n} m^{2} \sum_{a\not= b}^{n} g_{ab} \[G_{ab}\] \; \; .
\eq	

The correlation functions $G_{ab}(p)$ and the order parameter $m$ are then determined by 
the following saddle-point equations:

\bq
\lb{r19a}
\fr{\d f}{\d \tl{G}(p)} &=& 0
\\
\nn
\lb{r19}
\fr{\d f}{\d G_{ab}(p)} &=& 0 \; \; \; \; (a\not= b)
\\
\nn
\lb{r20}
\fr{\d f}{\d m} &=& 0 \; \; .
\eq
By using the explicit expression of the free energy density, eq.(\ref{r18}), 
one obtains

\bq
\lb{r21}
&& G_{ab}^{-1}(p) = (p^{2} +\t)\d_{ab} + g \[\tl{G}\] \d_{ab} + 2g_{ab} \[G_{ab}\] + 
                    g m^{2} \d_{ab} + 2 m^{2} g_{ab}
\\
\nn
\lb{r22}
&& m \( \t + g m^{2} + (3g-2u)\[\tl{G}\] + 
                    \fr{2}{n} \sum_{a\not= b}^{n} g_{ab} \[G_{ab}\] \) = 0 \; \; .
\eq
According to eq.(\ref{r21}) one finds that the trial correlation 
function has the following structure:

\be
\lb{r23}
G_{ab}^{-1}(p) = (p^{2} +\t)\d_{ab} + \mu_{ab} \; \; ,
\ee
where the matrix $\mu_{ab}$ is defined by 

\be
\lb{r24}
\mu_{ab} = \( g \[\tl{G}\]  + g m^{2} \) \; \d_{ab}
         + 2g_{ab} \[G_{ab}\] +  2 g_{ab} m^{2}   \; \; .
\ee

For finding explicit solutions of this equation one needs to make an assumption
about the replica structure of the matrix $\mu_{ab}$. In what follows we
assume that this matrix is replica symmetric; in Appendix A we give the 
formal proof that eq.(\ref{r24}) has no solutions with the Parisi replica 
symmetry breaking structure for the matrix $\mu_{ab}$. The replica symmetric ansatz 
implies that the matrix $\mu_{ab}$ is defined by only two parameters,

\be 
\lb{r25}
\mu_{ab} = (\tl{\mu} + \mu)\d_{ab} - \mu =
\left\{\ba \tl{\mu} \; \; \; ; \; \;  a=b 
           \\
          -\mu \; ; \; \;  a\not= b
       \ea
\right.
\ee
For the corresponding replica symmetric correlation function, 
defined by eq.(\ref{r23}), we find

\bq
\lb{r26}
G_{ab}(p; \lm, \mu) &=& \fr{1}{p^{2} + \lm} \; \d_{ab} + \fr{\mu}{(p^{2} + \lm)^{2}}
\nn
\nn
                    &\equiv& G_{c}(p;\lm) \; \d_{ab} + \mu \(G_{c}(p;\lm)\)^{2}
\eq
where the so-called "connected" part of the correlation function is given by

\be
\lb{r27}
G_{c}(p;\lm)  =  \fr{1}{p^{2} + \lm}
\ee
and instead of $\tl{\mu}$ (defined in eq.(\ref{r25})) we have introduced a physically
motivated "mass" parameter $\lm = \t + \tl{\mu} + \mu$ that determines the value of the 
correlation length ($R_{c} \sim \lm^{-1/2}$ in the present approximation). 

Note that according to eq.(\ref{r26}) the parameter $\mu$ is related to the value
of the spin-glass Edwards-Anderson (EA) order parameter, since

\be
\lb{r28}
q = \ol{\la\p\ra^{2}} = \lim_{n\to 0} \la\p_{a}(x)\p_{b}(x)\ra |_{(a\not= b)} = 
    \mu \[G_{c}^{2}\]  \; \; .
\ee
Thus, to be physically meaningful $\mu$ must be non-negative.

By using eqs.(\ref{r25}), (\ref{r26}) and (\ref{r7}), 
the corresponding saddle-point equations 
for the parameters $m$, $\lm$ and $\mu$ can be obtained 
from eqs.(\ref{r24}) and (\ref{r22}) as

\bq
\lb{r29}
&& \lm = \t + \mu + (3g-2u)\( \[G_{c}\] + \mu \[G_{c}^{2}\] \) + (3g-2u) m^{2}
\\
\nn
\lb{r30}
&& \mu = 2u\mu \[G_{c}^{2}\] + 2u m^{2}
\\
\nn
\lb{r31}
&& m\( \t + g m^{2} + (3g-2u)\( \[G_{c}\] + \mu\[G_{c}^{2}\] \) + 
   2 u \mu\[G_{c}^{2}\] \) = 0 \; \; .
\eq
The resulting the free energy density is then given by

\bq
\lb{r32}
f(m, \lm, \mu) &=& -\2 \[\ln(G_{c})\] + \2(\t-\lm) \(\[G_{c}\] + \mu \[G_{c}^{2}\]\)
       +\4 (3g-2u) \(\[G_{c}\] + \mu \[G_{c}^{2}\]\)^{2} + 
\nn
\nn
&& + \2 u\mu^{2} \[G_{c}^{2}\]  + \2 (3g-2u) m^{2} \(\[G_{c}\] + \mu \[G_{c}^{2}\]\) + 
u\mu m^{2} \[G_{c}^{2}\] +
\nn
\nn
&& + \2 \t m^{2} + \4 g m^{4}  \; \; .
\eq

\sectio{Phase diagram in $D=4$}

In this Section we study all possible solutions of the saddle-point equations 
(\ref{r29})-(\ref{r31}). As usual, it is assumed that 
the non-Gaussian coupling parameters $g$ and $u$
of the original replica Hamiltonian, eq.(\ref{r6}), are small: 
$g \ll 1 , \;  u \ll 1$.
Besides, we consider $u \sim g$, which qualitatively corresponds to the situation
of "finite disorder strength" since then the parameter $u$ describing 
the disorder strength of
is of the same order as the coupling parameter $g$ of the pure system.
As will be discussed in section 5, the analysis of the validity of the present (first-order)
Gaussian approximation shows that it may give reasonable results provided the 
ratio $u/g$ stays within certain numerical bounds (see Section 5). For the moment,
however, it is sufficient to assume that
$u < \fr{3}{2} g$ (see below). Finally, 
since we are only interested in the large-scale (continuous limit) 
properties of the system we consider the region of parameter space where
the mass $\lm$ of the connected correlation function
is also small: $\lm \ll 1$.

To simplify the algebra and for a qualitative presentation 
of the phase diagram 
it is convenient to consider first the solutions 
of the saddle-point equations in dimension $D=4$. 
Generalization of the results
for dimensions below four will be given in Section 5 
(it will also be shown that for $D > 4$
the spin-glass solution does not exist). 

For  $D = 4$ and for $\lm \ll 1$ one has

\bq
\lb{r33} 
\[G_{c}\] &=& \int_{|p|<1}\frac{d^{4}p}{(2\pi)^{4}} \fr{1}{p^{2} + \lm} \simeq
              C\(1 - \lm\ln(\fr{1}{\lm}) \)
\\
\nn
\lb{r34} 
\[G_{c}^{2}\] &=& \int_{|p|<1}\frac{d^{4}p}{(2\pi)^{4}} \fr{1}{(p^{2} + \lm)^{2}} \simeq
              C\(\ln(\fr{1}{\lm}) - 1\)
\eq
where $C = 1/16\pi^{2}$.

\subsection{Paramagnetic solution}

In the paramagnetic state, the ferromagnetic and spin-glass order parameters
are both zero ($m = \mu = 0$) and there is only one saddle-point equation (\ref{r29})
for the mass parameter $\lm$

\be
\lb{r35}
\lm = \t + (3g-2u)\[G_{c}\] \; \; .
\ee
Using eq.(\ref{r33}) leads to the following equation:

\be
\lb{r36}
\lm + C(3g -2u) \lm \ln\(\fr{1}{\lm}\) = \t +  C(3g -2u) 
\ee
which provides the dependence $\lm = \lm(\t)$. The solution of this equation
makes physical sense only for $\lm \geq 0$, and therefore this condition defines the 
limit of existence of the paramagnetic phase. Provided $3g > 2u$ the above equation
yields positive (physical) solutions for $\lm(\t)$ only for temperatures such that

\be
\lb{r37}
\t \geq \t_{c} = - C (3g - 2u) \; \; .
\ee
If one finds that the ferromagnetic solution appears just below $\t_{c}$,
then the temperature $\t = \t_{c}$ would correspond to the paramagnetic-ferromagnetic
phase transition point. However, it will be shown below that this is not the case.
In fact, below a certain temperature $\t_{sg} > \t_{c}$ a spin-glass solution
(with $\mu \not= 0$) appears, and at temperatures $\t < \t_{sg}$ it is the 
spin-glass state that turns out to be stable, while the paramagnetic one becomes
unstable.

\subsection{Spin-glass solution}

The spin-glass state is defined by two saddle-point equations (\ref{r29}) and (\ref{r30}):

\bq
\lb{r38}
 \lm &=& \t + \mu + (3g-2u)\( \[G_{c}\] + \mu \[G_{c}^{2}\] \)
\\
\nn
\lb{r39}
 \mu &=& 2u\mu \[G_{c}^{2}\]
\eq
which define two non-zero (positive) order parameters: $\mu(\t)$ and $\lm(\t)$.
From the last equation one immediately finds that for 
$\mu \not= 0$ the mass parameter $\lm$
becomes temperature independent, $\lm = \lm_{o}$, and the value of $\lm_{o}$ 
is defined by the condition

\be
\lb{r40}
\[G_{c}^{2}\] = \fr{1}{2u} \; \; .
\ee
Correspondingly, eq.(\ref{r38}) yields the following solution for the spin-glass 
order parameter:

\be
\lb{r41}
\mu(\t) = \fr{2u}{3g}\(\lm_{o} - (3g-2u)\[G_{c}\] - \t \) \; \; .
\ee
By making use of eqs.(\ref{r33}) and (\ref{r34}) one can find the solutions 
for $\lm_{o}$ and 
$\mu(\t)$ explicitly:

\bq
\lb{r42}
\lm_{o} &=& \exp\( -\fr{1}{2C u} - 1\)
\\
\nn
\lb{r43}
\mu(\t) &=&  \fr{2u}{3g}\(\t_{sg} - \t \)
\eq
where

\be
\lb{r44}
\t_{sg} = \lm_{o} -(3g-2u)\[G_{c}\] = \t_{c} + \( \fr{3g}{2u} + C(3g-2u) \) \lm_{o} \; \;  > \; \t_{c}
\ee
and $\t_{c}$ is the putative paramagnetic critical point discussed above.
The solution for $\mu > 0$ appears (i.e. becomes physical) only for $\t < \t_{sg}$, 
and therefore
the point $\t_{sg}$ can be associated with the spin-glass phase transition temperature. 
Note that according to eqs.(\ref{r44}) and (\ref{r35}), the value of $\lm$ 
in the paramagnetic phase at $\t = \t_{sg}$ is equal to $\lm_{o}$ 
(for $\t >\t_{sg}$, $\lm(\t) > \lm_{o}$ and for 
$\t < \t_{sg}$, $\lm(\t) < \lm_{o}$). 
Since $\mu(\t_{sg}) = 0$ whether one comes from the paramagnetic or 
the spin-glass phase, the transition into the spin-glass phase is clearly
continuous.

In Appendix B we present a detailed study of the
stability of the spin-glass and the paramagnetic solutions
obtained above. It is shown there that 
 for $\t > \t_{sg}$ the only stable state of the system is paramagnetic, 
while for $\t < \t_{sg}$
the paramagnetic solution becomes unstable and the stable state 
of the system is the spin-glass phase.

\subsection{Ferromagnetic solution}

We finally consider the ferromagnetic solution of 
the saddle-point equations (\ref{r29})-(\ref{r31})
in which all three parameters $\lm$, $\mu$ and $m$ are non-zero. 
After some simple algebra we find

\bq
\lb{r57}
m^{2} &=& \fr{1}{2g} \lm
\\
\nn
\lb{r58}
\mu &=& \fr{u}{g} \fr{\lm}{1 - 2 u \[G^{2}\]}
\eq
where the parameter $\lm(\t)$ is obtained from the following equation:

\be
\lb{r59}
\lm = \t + (3g-2u)\[G\] + \fr{3}{2} \fr{\lm}{1 - 2 u \[G^{2}\]} \; \; .
\ee
Substituting eqs.(\ref{r33}) and (\ref{r34}) into the above equation gives

\be
\lb{r60}
\lm \[\fr{3g}{2u} + C (3g - 2u) - C (3g - 2u) \ln\fr{\lm}{\lm_{o}} - 
      \fr{3}{4 C u} \fr{1}{\ln\fr{\lm}{\lm_{o}}} \] 
    \; = \; \t + C (3g - 2u) \; \; .
\ee
A simple analysis shows that upon lowering the temperature $\t$ 
a solution of this equation 
appears for the first time below a temperature $\t_{*}$.
This solution has a {\it finite} (non-zero)
value $\lm_{*}$ at $\t=\t_{*}$, which indicates that the phase 
transition into the ferromagnetic state
is first order. 
To the leading order in $g \ll 1$ and in $u \ll 1$ 
(and for $g/u \sim 1$) one finds

\be
\lb{r61}
\t_{*} \simeq -C (3g - 2u) - \fr{3}{4 C u} \exp\( - \fr{1}{2 C u} \) = 
              \t_{c} - \fr{3 e}{4 C u} \lm_{o} \; < \; \t_{c}
\ee
and

\be
\lb{r62}
\lm(\t=\t_{*}) \equiv \lm_{*} \simeq \exp\( - \fr{1}{2 C u} \) = e \lm_{o} \; \; .
\ee
By inserting the above value into eqs.(\ref{r57}) and (\ref{r58}), 
we find the corresponding 
values of the ferromagnetic and the spin-glass order parameters:

\bq
\lb{r63}
m^{2}_{*} &=& \fr{1}{2g} \lm_{*} \simeq \fr{1}{2g} \exp\( - \fr{1}{2 C u} \)
\\
\nn
\lb{r64}
\mu_{*} &\simeq& \fr{e}{2Cg} \lm_{o} = \fr{1}{2Cg} \exp\( - \fr{1}{2 C u} \) \; \; .
\eq
Straightforward calculations similar to those of Section 3.3 
show that the ferromagnetic
solution defined by eqs.(\ref{r57})-(\ref{r59}) 
is stable at all temperatures $\t < \t_{*}$.
Thus, below $\t_{*}$ both the spin-glass and the 
ferromagnetic solutions are (locally) stable
(this is the standard situation for first-order phase transitions). 
To determine 
which of these two states is the global minimum of the free energy
at a given temperature 
we have to compare the corresponding values of their free energies.

\subsection{First-order phase transition between spin-glass and ferromagnetic states}

Substituting the spin-glass solution, eqs.(\ref{r40})-(\ref{r44}), 
into eq.(\ref{r32}) provides
the value of the free energy density of the spin-glass state 
(in the leading order in $g,u \ll 1$):

\be
\lb{r65}
f_{sg}(t) \simeq f_{0}(\t) + \fr{1}{8}\lm_{o}^{2} - 
            \fr{C u}{3g} \lm_{o} (\t - \t_{c})
\ee
where

\be
\lb{r66}
f_{0}(\t) = \fr{u}{6g} C (3g-2u) + \fr{u}{3g} \t - \fr{1}{12} \t^{2} \; \; .
\ee
On the other hand,
for the ferromagnetic solution, eqs.(\ref{r57})-(\ref{r60}), one gets

\bq
\lb{r67}
f_{f}(\t) &\simeq& f_{0}(\t) +\fr{\lm}{6g} \( 1 + 2Cu \ln\lm \) (\t - \t_{c})
                  + \fr{C^{2}u(3g-2u)}{6g} \lm^{2} \ln^{2}\lm +
\nn
\nn
         && + C \(\fr{(3g-2u}{6g} - \4 \) \lm^{2} \ln\lm 
           + \fr{1}{24g} \lm^{2} - \fr{1}{8}\lm^{2}
\eq
where the value of $\lm(\t)$ is given by eq.(\ref{r60}). Let us redefine

\bq
\lb{r68}
\lm(\t) &\equiv& \lm_{o} x(\t)
\\
\nn
\lb{r69}
\t - \t_{c} &\equiv& -\lm_{o} t
\eq
By making the above change of variables
in the saddle-point equation (\ref{r60}), one obtains

\be
\lb{r70}
x \[\fr{3g}{2u} + C (3g - 2u) (1 - \ln x) - 
      \fr{3}{4 C u} \fr{1}{\ln x} \] \; = \; t \; \; .
\ee
Therefore, to the leading order in $u,g \ll 1$ the value 
of the parameter $x$ as the
function of the reduced temperature $t$ 
is defined by the following equation:

\be
\lb{r71}
\fr{3}{4 C u} \fr{x}{\ln x} \; \simeq \; t \; \; .
\ee
Assuming that at the point of the phase transition, 
i.e., when $f_{f} = f_{sg}$, the value of the 
parameter $x(t)$ is of order one, we find 
for the difference of the free energies,
eqs.(\ref{r65}),(\ref{r67}) (in the leading order 
in $u,g \ll 1$ and $\lm_{o} \ll 1$),

\be
\lb{r72}
f_{f} - f_{sg} \simeq \fr{C u}{3g} \lm_{o}^{2} t 
       \(1 - x - x\ln x\) + \fr{1}{8g} \lm_{o}^{2} x^{2} \; \; .
\ee
Thus, the transition point ($f_{f} - f_{sg} = 0$) 
is defined by the following equation: 

\be
\lb{r73}
\fr{C}{3} u t \(1 - x - x\ln x\) \; = \; \fr{1}{8} x^{2} \; \; .
\ee
Combining eqs.(\ref{r71}) and (\ref{r73}),
we finally derive the equation for the parameter $x$
at the phase transition point,

\be
\lb{r74}
\fr{3}{2} x - x \ln x = 1  \; \; .
\ee
This equation has a unique solution 
$x = x_{f} \sim 1$ ($x_{f} > 1$). Substituting 
$x_{f}$ into eqs.(\ref{r71}) and (\ref{r69}) one obtains the  
temperature of the (first-order)
phase transition between the spin-glass and the ferromagnetic phases:

\be
\lb{r75}
\t_{f} = \t_{c} - \fr{3}{4 C u} 
       \fr{x_{f}}{\ln x_{f}} \lm_{o} \; \; ,
\ee
which is less than $\t_{*}$.

\sectio{Singularities at the spin-glass phase transition}

Within the Gaussian variational approximation the (connected) correlation 
functions can be obtained by adding source terms to the 
replica Hamiltonian, eq.(\ref{r6}),
and by approximating the free energy functional to the first-order 
cumulant as in eq.(\ref{r12}).
This latter then generates the (connected) correlation functions 
that are obtained
by functional differentiation with respect to the source terms. 
By introducing source terms linearly coupled to the fields 
$\f_{a}(x)$, one derives
the usual correlation functions, whose expression (for the two-point
functions) coincides with that given in section 2. 
To study the singularity at the spin-glass
transition it is more convenient to introduce source terms linearly coupled 
to the composite  operators $\2 \f_{a}(x)\f_{b}(x)$. 
This leads us to the following replica Hamiltonian:

\be
\lb{r76}
H^{(n)}\[\f_{a}(x); {\bf{\D}}(x)\] = H^{(n)}\[\f_{a}(x); {\bf{0}}\] - 
        \2 \int d^{D}x \sum_{a,b=1}^{n} \D_{ab}(x) \f_{a}(x)\f_{b}(x)
\ee
where $H^{(n)}\[\f_{a}(x); {\bf{0}}\]$ is given by eq.(\ref{r6}) and,
because we are ultimately interested in the replica-symmetric solution, 
the source term is taken as

\be
\lb{r77}
\D_{ab}(x) = \(\tl{\D}(x) + \D(x)\) \d_{ab} - \D(x) \; \; .
\ee

Since we only study here the paramagnetic and the spin-glass phases, 
we can set $m=0$.
A Gaussian trial Hamiltonian is chosen as before, but the presence of 
space-dependent source terms breaks the translational invariance and requires 
trial Green functions that depend on two space points:

\be
\lb{r78}
H^{(n)}_{g}[\f_{a}|{\bf{G}}] = \2 \int\int d^{D}x d^{D}x' 
           \sum_{ab=1}^{n} \f_{a}(x) \[{\bf{G}}^{-1}\]^{xx'}_{ab} \f_{b}(x')
\ee
where

\be
\lb{r79}
\[{\bf{G}}^{-1}\]^{xx'}_{ab} = 
\[\(\fr{\partial^{2}}{\partial x \partial x'} + \t \) \d_{ab} + \mu_{ab}(x) \] 
                 \d (x-x') \; \; ,
\ee
which simply generalizes eq.(\ref{r23}).  For the replica-symmetric solution, 
the variational parameter $\mu_{ab}(x)$ can be written as 

\be
\lb{r80}
\mu_{ab}(x) = \(\tl{\mu}(x) + \mu(x)\) \d_{ab} - \mu(x) \; \; ,
\ee
whereas the Green functions can be written as

\be
\lb{r81}
G_{ab}^{xx'} = G_{c}^{xx'} \d_{ab} + G_{d}^{xx'}
\ee
where $G_{c}$ and $G_{d}$ represent, 
as usual in the presence of quenched disorder,
the "connected" and "disconnected" parts and can be expressed in terms of $\t$,
$\tl{\mu}(x)$ and $\mu(x)$ by inverting eq.(\ref{r79}).

One can then follow the procedure used in section 2: 
the first-order cumulant approximation in 
the deviation $(H^{(n)} - H^{(n)}_{g})$ 
provides an upper bound for the free energy 
functional, and minimizing with respect to the trial Green function
elements leads to saddle-point equations that in the limit
$n\to 0$ for a replica-symmetric scheme reduce to

\bq
\lb{r82}
\tl{\mu}(x) + \tl{\D}(x) &=& (3g-2u) \(G_{c}^{xx} + G_{d}^{xx} \)
\\
\nn
\lb{r83}
\mu(x) + \D(x) &=& 2u G_{d}^{xx} \; \; .
\eq

When considered at the saddle-point characterized by the above equations, 
the variational free energy $F\[\tl{\D}(x), \D(x)\]$ can be used as 
generating functional for the correlation functions of the 
composite operators $\2 \f_{a}(x)\f_{b}(x)$. More precisely, one has

\bq
\lb{r84}
\fr{\d F}{\d\tl{\D}(x)} &=& -\2 \lim_{n\to 0} \la\f_{a}(x)\f_{a}(x)\ra =
                            -\2 \( G_{c}^{xx} + G_{d}^{xx} \)
\\
\nn
\lb{r85}
\fr{\d F}{\d \D(x)} &=& \left.
            \2 \lim_{n\to 0} \la\f_{a}(x)\f_{b}(x)\ra \right|_{(a\not= b)} =
                            \2 G_{d}^{xx} 
\eq
In the limit $\tl{\D} = \D = 0$ the above equations 
reduce to the expressions
already given in section 2, in particular 
eq.(\ref{r85}) reduces to 
eq.(\ref{r28}) for the spin-glass order parameter: 
indeed, in the absence of 
source terms translational invariance is recovered and 
$G_{d}^{xx} = \[G_{d}(p)\] = \mu \[G_{c}^{2}(p)\]$.

The second functional derivatives of $F$ provide the wanted two-points,
four-field correlation functions, namely:

\bq
\lb{r86}
\fr{\d^{2} F}{\d\tl{\D}(x)\d\tl{\D}(x')} &=& -\4 \lim_{n\to 0} 
  \fr{1}{n} \sum_{a,b=1}^{n}\la\f_{a}^{2}(x)\f_{b}^{2}(x')\ra_{c}
\\
\nn
\lb{r87}
\fr{\d^{2} F}{\d\tl{\D}(x)\d\D(x')} &=& \4 \lim_{n\to 0} 
    \fr{1}{n} \sum_{a,b=1}^{n}\la\f_{a}^{2}(x) 
    \(\sum_{c\not= b}^{n} \f_{c}(x')\f_{b}(x')\) \ra_{c}
\\
\nn
\lb{r88}
\fr{\d^{2} F}{\d\D(x)\d\D(x')} &=& -\4 \lim_{n\to 0} \fr{1}{n} 
  \sum_{a,b=1}^{n}\la \(\sum_{c\not= a}^{n} \f_{a}(x)\f_{c}(x)\)
                      \(\sum_{d\not= b}^{n} \f_{b}(x')\f_{d}(x')\) \ra_{c}
\eq
where $\la(...)\ra_{c}$ denotes a cumulant average for the composite operators.

We are interested in the case $\tl{\D} = \D = 0$, 
case in which translational 
invariance is recovered and the two-point functions are diagonal in 
momentum space. We then define the $2\times 2$ 
momentum-dependent susceptibility 
matrix $\mbox{\boldmath $\chi$}(p)$ by 

\bq
\lb{r89}
\chi_{11}(p) &=& \left.
             \fr{\d^{2} F}{\d\tl{\D}(-p)\d\tl{\D}(p)}\right|_{\tl{\D}=\D=0} \; ; \; \; 
\chi_{12}(p)  = \left.
             \fr{\d^{2} F}{\d\tl{\D}(-p)\d\D(p)}\right|_{\tl{\D}=\D=0} ;
\nn
\nn
\chi_{21}(p) &=& \left.
             \fr{\d^{2} F}{\d\D(-p)\d\tl{\D}(p)}\right|_{\tl{\D}=\D=0} \; ; \; \; 
\chi_{22}(p) = \left.
             \fr{\d^{2} F}{\d\D(-p)\d\D(p)}\right|_{\tl{\D}=\D=0} \; \; .
\eq
The expression of these $p$-dependent susceptibilities in terms of the
correlation functions follow from eqs.(\ref{r86})-(\ref{r88}). By combining the
above definitions, the expression of the free energy functional at the first-order
cumulant approximation and the saddle-point equations, eq.(\ref{r82})-(\ref{r83}),
one obtains after some algebra the following expressions for the $p$-dependent
susceptibility matrix:

\be
\lb{r90}
\mbox{\boldmath $\chi$}(p) = \2 \( \; 
\begin{tabular}{cc}
$\fr{1}{(3g-2u)}$ & $0$ \\
\\
$0$ & $\fr{1}{2u}$ \\
\end{tabular}
\; \) 
\( {\bf{M}}^{-1}(p) - {\bf{I}} \)
\ee
where ${\bf{I}}$ is the identity matrix and the $p$-dependent matrix 
${\bf{M}}(p)$ is defined by 

\be
\lb{r91}
{\bf{M}}(p) = \( \; 
\begin{tabular}{cc}
$1 + (3g-2u) \(I(p) + 2\mu J(p)\)$ & $2(3g-2u) \mu J(p)$ \\         
\\
 $4u\mu J(p)$  & $ 1 - 2u \( I(p) - 2\mu J(p)\)$\\
\end{tabular}
\; \)
\ee
with 

\bq
\lb{r92}
I(p) &=& \int_{|p|<1}\fr{d^{D}k}{(2\pi)^{D}} 
       \fr{1}{({\bf{k}}^{2} + \lm)({\bf{(k + p)}}^{2} + \lm)}
\\
\nn
\lb{r93}
J(p) &=& -\2 \fr{\partial I(p)}{\partial \lm} =  
       \int_{|p|<1}\fr{d^{D}k}{(2\pi)^{D}} 
       \fr{1}{({\bf{k}}^{2} + \lm)^{2}({\bf{(k + p)}}^{2} + \lm)} \; \; ,
\eq
and $\lm$ and $\mu$ are given by eqs.(\ref{r38}) and (\ref{r39}), respectively.
From the above equations one immediately obtains the expression for the 
susceptibility matrix in the paramagnetic phase (where $\mu = 0$):

\be
\lb{r94}
\mbox{\boldmath $\chi$}(p) = \2 \( \; 
\begin{tabular}{cc}
$-\fr{I(p)}{1 + (3g-2u) I(p)}$ & $0$ \\
\\
$0$ & $\fr{I(p)}{1-2u I(p)}$ \\
\end{tabular}
\; \)  \; \; ,
\ee
from which one derives that the susceptibility 
$\chi_{22}(p=0) = \(\partial^{2}F/\partial \D^{2}\)|_{(\tl{\D}=\D=0)}$
diverges when $1-2u I(0) = 0$, i.e., by using eq.(\ref{r92}),
when $\[G_{c}^{2}\] = 1/2u$. 
This point is precisely attained at the transition 
to the spin-glass state, $\t = \t_{sg}$, where, for $D=4$, 
$\t_{sg}$ is given by eq.(\ref{r44})
($\lm$ is then equal to $\lm_{o}$ given by eq.(\ref{r42})).
To derive the critical behavior of $\chi_{22}(p)$ when approaching the 
spin-glass transition from above, we use the small-$p$ expansion
of $I(p)$ in $D=4$,

\be
\lb{r95}
I(p) \simeq C \(\ln\(\fr{1}{\lm}\) - 1 \) - \fr{C_{2}}{\lm} p^{2}
\ee
where $C_{2} > 0$. After defining $\lm = \lm_{o} + \d \lm$, 
$\d \lm \to 0^{+}$, this gives

\be
\lb{r96}
\chi_{22}(p) \simeq \(\fr{\lm_{o}}{8 u^{2} C_{2}}\) \; 
                    \fr{1}{ p^{2} + \fr{C}{C_{2}} \d \lm} \; \; .
\ee
Using eqs.(\ref{r36}), (\ref{r42}) and (\ref{r44}) one finds that when
$\t \to \t_{sg}^{+}$,

\be
\lb{r97}
\d \lm \simeq \fr{2u}{3g} (\t - \t_{sg}) \; \; ,
\ee
which finally leads to

\be
\lb{r98}
\chi_{22}(p) \simeq \(\fr{\lm_{o}}{8 u^{2} C_{2}}\) \; 
       \fr{1}{ p^{2} + \fr{2u C}{3g C_{2}} (\t - \t_{sg})} \; \; .
\ee

Consider now the spin-glass phase (in $D=4$).
One then has $2u I(0) = 1$, $\lm = \lm_{o}$,
and $\mu = \fr{2u}{3g} (\t_{sg} - \t)$, so that for small $p$,

\bq
\lb{r99}
I(p) &\simeq& \fr{1}{2u} - \fr{C_{2}}{\lm_{o}} p^{2}
\\
\nn
\lb{r100}
J(p) &\simeq& \fr{C}{2\lm_{o}} - \fr{C_{2}}{2\lm_{o}} p^{2} \; \; .
\eq
The determinant of the matrix ${\bf{M}}(p)$, eq.(\ref{r91}), 
can now be expressed at the leading order as

\be
\lb{r101}
det({\bf{M}}(p)) \simeq \fr{3gC_{2}}{\lm_{o}}
          \( p^{2} + \fr{2u C}{3g C_{2}} (\t_{sg} - \t) \)
\ee
so that when $\t \to \t_{sg}^{-}$, the susceptibilities $\chi_{11}(p)$ 
stay $\chi_{12}(p)$ are finite, whereas

\be
\lb{r102}
\chi_{22}(p) \simeq \(\fr{\lm_{o}}{8 u^{2} C_{2}}\) \; 
            \fr{1}{ p^{2} + \fr{2u C}{3g C_{2}} (\t_{sg} - \t) } \; \; .
\ee

One concludes from the above formulas that the susceptibility 
associated with the external field $\D$ that couples to the 
Edwards-Anderson order parameter $q$, i.e.,
$\chi_{22}(p) = \fr{\d^{2} F}{\d\D(-p)\d\D(p)}$, 
diverges when the critical point $\t_{sg}$ is approached both from
above and below as $|\t - \t_{sg}|^{-1}$ (in $D=4$), whereas the 
associated correlation length
(that characterizes the long-distance behavior of the two-point
composite-field correlation function, eq.(\ref{r88}))
diverges as $|\t - \t_{sg}|^{-1/2}$ (in $D=4$). The corresponding 
critical exponents, $\g = 1$, $\nu = 1/2$, are thus classical.

\sectio{Discussion}

\subsection{Dimensions other than $D=4$}

The situation  in dimensions $D > 4$ is quite simple. 
There, the value of the integral

\be
\lb{r112}
\[G_{c}^{2}\] \equiv \Ip \fr{1}{p^{2} + \lm}   
\ee
remains finite (not diverging) in the limit $\lm \to 0$. 
Therefore when $u \ll 1$ the only solution of the saddle-point equation
(\ref{r39}) for the spin-glass order parameter
is trivial, $\mu = 0$, so that there is no spin-glass solution
in dimensions $D > 4$. Thus, one recovers in this case the
standard scenario: upon lowering the temperature, the only 
phase transition that takes place in the system
is a second-order phase transition from the paramagnetic 
to the ferromagnetic phase,
and this phase transition is described by the Gaussian theory.

\vspace{5mm}

On the other hand, the phase diagram in dimensions $D < 4$
turns out to be similar to that in $D = 4$ at a qualitative level.
The integral, eq.(\ref{r112}), diverges
in the limit $\lm \to 0$, and, therefore, when $u \ll 1$
there is always a spin-glass solution $\mu \not= 0$
of the saddle-point equation (\ref{r39}).
Thus, in this case  one recovers {\it within the Gaussian variational
approximation} a phase diagram
similar to that in dimension $D = 4$, where the 
paramagnetic phase is separated from the ferromagnetic
one by an intermediate spin-glass phase.

\subsection {Validity of the Gaussian variational approximation}

Since the main result of the present study, namely, the existence
of a spin-glass phase separating the paramagnetic and ferromagnetic phases,
is in apparent contradiction with the generally accepted view
on the phase diagram of the disordered Ising ferromagnet,
the limits of validity of the Gaussian variational approximation
used in this paper requires a detailed study.

It is well known that the Gaussian variational approach becomes
exact when the number of spin components tends to infinity
(see footnote on page 5).
Otherwise (in particular, the Ising model is very far from this limit)
it is not more than an approximation characterized by certain
bounds of validity (if any), and ignoring these bounds may just lead
to wrong conclusions. The typical example is the pure Ising
model in dimensions $D \leq 4$: there, one can easily check
(using eqs.(\ref{r29}) and (\ref{r31}) with $u=\mu=0$) that according 
to the Gaussian variational approximation the transition from the 
paramagnetic to the ferromagnetic phase turns out to be first-order,
which is of course incorrect.

The validity of the first-order cumulant approximation in the 
deviation $(H^{(n)} - H^{(n)}_{g})$, described in section 2,
can be checked by estimating the contribution from the 
higher order terms. In the vicinity of the critical
temperature $T_{c}$ of the (supposed) paramagnetic-ferromagnetic 
phase transition, for $|T - T_{c}|/T_{c} \equiv \t \ll 1$,
a systematic account of these contributions
can be done in terms of the usual renormalization group (RG) procedure,
which yields an effective scale dependence of the renormalized
non-Gaussian interaction parameters $g$ and $u$. 
In dimensions $D=4$,
in the so-called one-loop approximation, the scale evolution
of $g$ and $u$ is described by the following well-known
RG equations (see e.g. \cite{crit1}, \cite{rg-rsb}, \cite{intro}):

\bq
\lb{r103}
\fr{d}{d\xi} g(\xi) = -6 C \(3g - 4u\) g 
                    + O\(g^{3}; g^{2}u; g u^{2}; u^{3}\)
\\
\nn
\lb{r104}
\fr{d}{d\xi} u(\xi) = -4 C \(3g - 4u\) u 
                    + O\(g^{3}; g^{2}u; g u^{2}; u^{3}\)
\eq
where, as usual, $C = 1/16\pi^{2}$, and $\xi \equiv \ln L$ is the 
standard RG rescaling parameter which is equal to the logarithm of the
spatial scale. According to these equations one can conclude that the
renormalization of
the parameters $g$ and $u$ remains irrelevant (so that the higher order
terms of the perturbation theory are not important) and that the theory
remains Gaussian at scales bounded by the conditions

\bq
\lb{r105}
6 C |3g_{0} - 4u_{0}| g_{0} \xi \ll g_{0}
\\
\nn
\lb{r106}
4 C |3g_{0} - 4u_{0}| u_{0} \xi \ll u_{0} 
\eq 
where $g_{0} \equiv g(\xi=0)$ and $u_{0} \equiv u(\xi=0)$ are
the "bare" (microscopic) values of the parameters $g$ and $u$.
These two conditions are satisfied when

\be
\lb{r107}
\xi \ll \xi_{*} \sim \fr{1}{6 C |3g_{0} - 4u_{0}|} \; \; ,
\ee
or, in terms of the spatial scale $L$, until

\be
\lb{r108}
L \ll L_{*} \sim \exp\{ \fr{1}{6 C |3g_{0} - 4u_{0}|} \} \; \; .
\ee
Since the temperature and the spatial scales are related by
$|\t| \sim L^{-2}$, one finds that the higher-order non-Gaussian
corrections remain irrelevant only at temperatures not too close
to the (supposed) paramagnetic-ferromagnetic critical point:

\be
\lb{r109}
|\t| \gg \t_{*} \sim \exp\{ - \fr{1}{3 C |3g_{0} - 4u_{0}|} \} \; \; .
\ee
Note that in the special case $u_{0} = \fr{3}{4} g_{0}$, the higher-order
terms of the RG eqs.(\ref{r103})-(\ref{r104}) come into play,
and instead of the condition (\ref{r109}) one eventually obtains:
$\t_{*} \sim \exp\{ -(const)/g_{0}^{2}\}$. Note also that in the case
of the pure system, $u_{0} \equiv 0$, eq.(\ref{r109}) yields the
well known temperature scale $\t_{*} \sim \exp\{ - 16\pi^{2}/9g_{0}\}$,
such that when $|\t| \gg \t_{*}$ the behavior of the system is 
effectively Gaussian, while in the close vicinity of the critical point, 
for $|\t| \ll \t_{*}$, the non-Gaussian fluctuations become dominant;
in particular, this shows that the first-order Gaussian 
cumulant approximation breaks down for $|\t| \ll \t_{*}$
(it is at this temperature scale that this approximation wrongly predicts
the first-order nature of the phase transition).

According to the calculations of section 3, the spin-glass phase
separating the paramagnetic and ferromagnetic phases exists in
a temperature interval of order of $\lm_{o} \simeq \exp\{-\fr{1}{2 u C}\}$
around $\T$ (where $\T$ is the critical temperature of the putative
paramagnetic-ferromagnetic phase transition). Since the approximation
used in the present approach is valid only outside the temperature
given by eq.(\ref{r109}), we need this "dangerous"
temperature interval to be well inside the spin-glass phase, i.e.,

\be
\lb{r110}
\exp\{ - \fr{1}{3 C |3g_{0} - 4u_{0}|} \} \ll \exp\{-\fr{1}{2 u C}\} \; \; .
\ee
If the above condition is satisfied,
 the spin-glass state 
(obtained in terms of the present Gaussian variational approximation)
would set in when lowering the temperature
well before the non-Gaussian critical fluctuations
of the fields $\f_{a}(x)$ (given by the higher-order terms of the
perturbation theory) become relevant. 

It should be stressed, however, that this does not
guarantee that the critical behavior at the spin-glass phase transition
itself is correctly described by the Gaussian theory (as it is derived
in section 4), since for an adequate description  of this phase
transition one needs to take into account critical fluctuations
of the spin-glass order parameter $q_{ab}(x)= \f_{a}(x)\f_{b}(x)$
and not just those of the fields $\f_{a}(x)$. 
Actually, and although less likely, one can not exclude that the
fluctuations of the composite field $q_{ab}(x)$ will even destroy
the spin-glass phase itself. This question is left for future
investigations.

In addition, when further lowering the temperature, the 
transition point from the spin-glass to the ferromagnetic one
is also separated from $\T$ by an interval of the
order of $\lm_{o}$. Therefore, provided the condition (\ref{r110})
is satisfied, one can conclude that the first-order nature
of this phase transition corresponds  indeed to the proper physical phenomenon,
and is not just an artifact of the method used.

Since all the calculations of the present study are done
for $g \ll 1$ and $u \ll 1$, one can easily see that the 
condition (\ref{r110}) is satisfied provided

\be
\lb{r111}
\fr{10}{9}  < \fr{g_{0}}{u_{0}} < \fr{14}{9} \; \; ,
\ee
so that the parameter $u$ describing the disorder strength must be
of the same order as the interaction parameter $g$ 
of the pure system. It is in this sense that we characterize such system as
having a {\it finite} strength of disorder (although both $u$ and $g$
are kept small). 
 
Thus, we have found at least a limited region of the parameters, 
defined by eq.(\ref{r111}), 
where the approximation used in this paper appears to be reasonable and for 
which an intermediate spin-glass phase could exist. Presumably,
it could exist also in a wider region (in particular for large
values of $u$ compared to $g$), where, however, it  
cannot be studied by the present method. Unfortunately,
in the framework of the present investigation we cannot
answer the question of whether the (exponentially narrow) 
intermediate spin-glass phase continues to exist in the limit 
$u \to 0$  or whether there exists a critical value $u_{c}$ such that
the spin-glass phase appears only for  $u  > u_{c}$.

\vspace{5mm}

At a qualitative level, the situation in dimensions $D = 4 - \e$ ($\e \ll 1$)
turns out to be similar to that in $D = 4$ (see Appendix C):
we find again the same restriction on the parameters $g$ and $u$ as in
eq.(\ref{r111}), but
in addition we obtain that the value of the parameter
$\e = 4 - D$ must remain small.
This shows, on one hand,  that the results
obtained in the present study are not unique to the dimension $D=4$
and can be continued down to lower dimensions; on the other 
hand, due to the restriction $\e \ll 1$, we cannot guarantee that
they would survive down to the physical dimension $D=3$ 
(where, for sure, the present approximation does not work).

\subsection{Conclusion}

The existence of a spin-glass phase in a "dilute ferromagnet" is 
{\it a priori} a puzzle \cite{Ma-R},\cite{DS}. 
In the absence of competing 
antiferromagnetic interactions and of the associated frustration,
how can the system ever be found in a state with frozen magnetizations
but without global magnetization?\footnote{
The fact that at and around the spin-glass transition, the 4-spin
spin-glass susceptibility has to be larger than the square of the 
2-spin ferromagnetic susceptibility may not be as severe a problem:
the presence of a large number of "ferromagnetic islands" and clusters
invalidates the inequalities obtained via a perturbative analysis \cite{DS}}
Indeed, for a spin-glass phase as 
predicted  by the Gaussian variational 
approximation to exist in a disordered Ising models, one
must have

(i) non-zero "frozen" local magnetizations in an extensive part
of the volume, so that the Edwards-Anderson order parameter
is non-zero;

(ii) a total magnetization equals to zero.

Truly frozen local magnetizations require symmetry breaking and can 
exist only on a  percolating cluster that diverges in the thermodynamic limit.
If $L$ is the linear size of the whole system (ultimately, $L \to \infty$),
the size of clusters with non-zero magnetization must scale as $L^{d}$
with $D \geq d  > 1$ (for Ising spins). On the other hand, 
on each cluster the magnetization can be either positive or negative.
One possible scenario to explain the occurrence of the spin-glass phase
with zero total magnetization is thus as follows.

We envisage that there exist a global temperature at wchich 
a large number $M$ of percolating clusters
of fractal dimension $d<D$ acquire a non-zero magnetization
(when $L\to\infty$), while being essentially decoupled from
each other. These clusters are formed from the "islands"
characterized by a predominantly
negative value of the local temperature $(\t - \d\t)$
and are 
separated from each other by regions of high local temperature.
In zero external magnetic field\footnote{
If one imposes a small external magnetic field, all cluster
magnetizations will have the same sign, i.e., the sign of the 
external field. Then, in the limit of vanishingly small field, the
overall spontaneous magnetization does not vanish. The situation, however,
is different if one considers from the beginning the case without
any applied magnetic field.}
the sign of the magnetization on each cluster is then randomly
determined (and subsequently frozen when the temperature  is decreased).
Since $M$ is large, there will be essentially (up to a relative factor of
order $1/\sqrt{M}$) as many positive as negative clusters.
So if $M$ scales with $L$ as $M \sim L^{D-d}$ (and $d < D$), the total 
magnetization, averaged over all clusters, is zero whereas the 
Edwards-Anderson order parameter is non-zero: the system 
is  then in a spin-glass phase.
It must be stressed that the appearance of the large number of percolating 
clusters can not be envisaged as a purely geometric,
disorder-controlled phenomenon.
thermal fluctuations come into play for helping the islands
to coalesce. This is necessary in order to have $M$ scale
as $L^{D-d}$ since in usual percolation the number of 
incipient spanning clusters at the percolation threshold can be 
larger but in $D < 6$ does not diverge with $L$ as a power law
\cite{perc}It must be stressed that the appearance of the large number of percolating 
clusters can not be envisaged as a purely geometric,
disorder-controlled phenomenon.
thermal fluctuations come into play for helping the islands
to coalesce. This is necessary in order to have $M$ scale
as $L^{D-d}$ since in usual percolation the number of 
incipient spanning clusters at the percolation threshold can be 
larger but in $D < 6$ does not diverge with $L$ as a power law
\cite{perc}

To settle the question of the existence of a spin-glass phase in the 
random temperature Ising ferromagnet, especially in $D=3$, further
investigation is clearly needed. A potential line of investigation
is a renormalization group analysis that treats on equal footing the 
primary fields $\f_{a}(x)$ and the composite fields 
$\f_{a}(x)\f_{b}(x)$ that are needed to describe the spin-glass phase.
This would clarify the problem of the upper critical dimension of the 
paramagnetic to spin-glass phase transition (recall that in our study
$D=4$ is the critical dimension above which the spin-glass phase
ceases to exist, it is not necessarily the upper critical dimension).
An interesting problem that is worth pursuing is the connection
to the phase behavior of the random field Ising model.
It has been suggested \cite{rf}, \cite{bd} that this latter system
possesses an intermediate spin-glass-like phase, and that a potential
source for the presence of this phase is the generation
in higher orders of perturbation theory of
"attractive" terms of the type $u \f_{a}^{2}\f_{b}^{2}$
\cite{bd}, i.e., terms similar to those present in the 
random temperature replica Hamiltonian. Although the 
kind of symmetry breaking leading to the spin-glass phase
in the random temperature Ising model is not possible in the 
random field models (within the replica formulation, 
random fields generate a nonzero field $\D$ that is linearly
coupled to the composite operator $\f_{a}(x)\f_{b}(x)$),
a comparative study may prove fruitful. Finally,
one should stress that 
absence of the replica symmetry breaking found within
the present approach could turn out to be 
an artifact of the Gaussian variational approximation because 
the corresponding saddle-point equations (see Appendix A) 
are in fact rather "fragile" with respect to the introduction of 
higher-order terms of the perturbation theory.

\vspace{15mm}

{\bf Acknowledgments}.

The authors are grateful to E. Br\'ezin and C. De Dominicis   
for stimulating discussions.

\vspace{15mm}

\sectio{Appendix A}

Consider again the saddle-point equations (\ref{r24})
in the spin-glass state ($m=0$):

\be
\lb{a1}
\mu_{ab} = g \[\tl{G}\] \d_{ab}   + 2g_{ab} \[G_{ab}\]
\ee
where

\be
\lb{a2}
G_{ab}^{-1}(p) = (p^{2} + \t) \d_{ab} + \mu_{ab}
\ee
and $\tl{G}(p) \equiv G_{a=b}(p)$.

\subsection{Continuous RSB}

In the case of a Parisi-like continuous RSB 
in the limit $n \to 0$ the matrices 
$G_{ab}$ and $\mu_{ab}$ are parametrized by their diagonal elements 
$\tl{G}$, $\tl{\mu}$ and by off-diagonal {\it functions} 
$G(p; x)$, $\mu(p; x)$ 
defined on the interval $x \in [0, 1]$ \cite{sg}:

\bq
\lb{a3}
G_{ab}(p) \; &\to& \; \(\tl{G}(p); \; G(p; x) \)
\\
\nn
\lb{a4}
\mu_{ab}(p) \; &\to& \; \(\tl{\mu}(p); \; \mu(p; x) \)
\eq
By using the Parisi algebra for inverting hierarchical matrices
\cite{gva1} one derives from eq.(\ref{a2})

\be
\lb{a5}
G(p; x) = - \fr{\mu(0)}{(p^{2} + \t + \tl{\mu} - \ol{\mu})^{2}} -
         \int_{0}^{x} dy \fr{\mu'(y)}{(p^{2} + \lm(y))^{2}}
\ee
where

\bq
\lb{a6}
\lm(y) &\equiv& \t + \tl{\mu} - \ol{\mu} - y \mu(y) 
           + \int_{0}^{y} dz \mu(z)
\\
\nn
\lb{a7}
\ol{\mu} &\equiv& \int_{0}^{1} dx \mu(x)
\\
\nn
\lb{a8}
\mu(0) &\equiv& \mu(x=0) \; \; .
\eq
Substituting eq.(\ref{a5}) into eq.(\ref{a1}) (for $a\not= b$), we find
the following equation for the unknown function $\mu(x)$:

\be
\lb{a9}
\mu(x) = 2u \Ip \fr{\mu(0)}{(p^{2} + \t + \tl{\mu} - \ol{\mu})^{2}} +
         2u \Ip \int_{0}^{x} dy \fr{\mu'(y)}{(p^{2} + \lm(y))^{2}} \; \; .
\ee
Differentiating both sides of this equation 
with respect to $x$ then leads to

\be
\lb{a10}
\mu'(x) \[ 2u \Ip \fr{1}{(p^{2} + \lm(x))^{2}} - 1 \] = 0 \; \; .
\ee
Thus, either one has $\mu'(x) = 0$, i.e.,  $\mu(x) = const$ 
(independent of $x$), or

\be
\lb{a11}
2u \Ip \fr{1}{(p^{2} + \lm(x))^{2}} - 1  = 0
\ee
which also yields $\mu(x) = const$. In both cases the solution is 
replica-symmetric (or step-like).
This proves that the saddle-point equation (\ref{a1}) 
cannot have solutions with continuous RSB. 

Nevertheless, the above proof does not exclude the possibility of 
other types of solutions with step-like RSB.

\subsection{One-step RSB}

We consider now a one-step RSB ansatz for the replica matrix $G_{ab}(p)$:
it is defined in terms of one diagonal element $\tl{g}(p)$, {\it two} 
off-diagonal elements $g_{1}(p)$ and $g_{0}(p)$, and the coordinate 
of the step $x_{o}$:

\be 
\lb{a12}
G_{ab}(p) = 
\left\{\ba \tl{g}(p) \; \; ; \mbox{for}  \;  a=b 
           \\
           \\
            g_{1}(p) \; \; ; \mbox{for} \; I\(\fr{a}{x_{o}}\) = I\(\fr{b}{x_{o}}\)
           \\
           \\
            g_{0}(p) \; \; ; \mbox{for} \; I\(\fr{a}{x_{o}}\) \not= I\(\fr{b}{x_{o}}\)
           \ea
\right.
\ee
where $I(x)$ is the integer valued function which is equal to the smallest integer
larger than or equal to $x$. Substituting the above equation into the general
expression, eq.(\ref{r18}), (with $m=0$),
one finds the following explicit expression for the free energy density:

\bq
\lb{a13}
F\[\tl{g}(p); g_{1}(p); g_{0}(p); x_{o}\] &=& 
  -\fr{1}{2x_{o}}\[\ln\(\tl{g} - g_{1} + x_{o}(g_{1}-g_{0}) \) \]
  -\2 \(1 - \fr{1}{x_{o}}\) \[\ln\(\tl{g} - g_{1}\) \] -
\nn
\nn
&&  -\[\fr{g_{0}}{2\(\tl{g} - g_{1} + x_{o}(g_{1}-g_{0}) \)} \]
    +\2 \[ (p^{2} + \t)\tl{g}\] +
\nn
\nn
&&  + \4 (3g-2u) \[\tl{g}\]^{2} + \2 u (1-x_{o})\[g_{1}\]^{2} 
    + \2 u x_{o}\[g_{0}\]^{2}
\eq
where, as usual, the symbol $\[(...)\]$ denotes the integration over $p$,
see eq.(\ref{r17a}). Variation of this free energy density with respect to the 
trial functions $g_{0}(p)$, $g_{1}(p)$ and $\tl{g}(p)$
yields three saddle-point equations,

\be
\lb{a14}
\fr{g_{0}(p)}{\(\tl{g}(p) - g_{1}(p) + x_{o}
              (g_{1}(p) - g_{0}(p)) \)^{2}} = 2 u \[g_{0}\]
\ee
\be
\lb{a15}
\fr{g_{1}(p) - g_{0}(p)}{\(\tl{g}(p) - g_{1}(p)\) 
\(\tl{g}(p) - g_{1}(p) + x_{o}(g_{1}(p) - g_{0}(p)) \)} = 
                2 u \[(g_{1} - g_{0})\]
\ee
\be
\lb{a16}
\fr{1}{\tl{g}(p) - g_{1}(p)} = p^{2} + \t + 
                    (3g - 2u)\[\tl{g}\] + 2 u \[g_{1}\] \; \; .
\ee
The last equation can be rewritten as follows,

\be 
\lb{a17}
\tl{g}(p) - g_{1}(p) = \fr{1}{p^{2} + 
                 \lm_{1}} \equiv G_{c}(p;\lm_{1}) \; \; ,
\ee
where the unknown parameter $\lm_{1}$ is defined by 

\be
\lb{a18}
\lm_{1} = \t + (3g - 2u)\[G_{c}\] + 3g \[g_{1}\] \; \; .
\ee
From eqs.(\ref{a15}) and (\ref{a14}) one finds

\bq
\lb{a19}
g_{1}(p) - g_{0}(p) &=& q_{1} G_{c}(p;\lm_{1})G_{c}\(p; (\lm_{1} - q_{1}x_{o})\)
\\
\nn
\lb{a20}
g_{0}(p) &=& q_{0} G_{c}^{2}\(p; (\lm_{1} - q_{1}x_{o})\) \; \; ,
\eq
and the unknown parameters $q_{1}$ and $q_{0}$ are defined by

\bq
\lb{a21}
\[G_{c}(\lm_{1}) G_{c}(\lm_{1} - q_{1}x_{o}) \] &=& \fr{1}{2u}
\\
\nn
\lb{a22}
q_{0} \(2u\[G_{c}^{2}(\lm_{1} - q_{1}x_{o})\] - 1 \) &=& 0 \; \; .
\eq
One more saddle-point equation is obtained 
by taking the derivative of the
free energy, eq.(\ref{a13}), with respect 
to the parameter $x_{o}$; it reads

\be
\lb{a23}
\[\ln G_{c}(\lm_{1})\] - \[\ln G_{c}(\lm_{1} - q_{1}x_{o})\] + 
q_{1}x_{o}\[G_{c}(\lm_{1} - q_{1}x_{o})\] =
\fr{1}{4u} (q_{1}x_{o})^{2}  \; \; .
\ee
In this way, we have obtained four equations, 
(\ref{a18}), (\ref{a21})-(\ref{a23}), for
four unknown parameters, $\lm_{1}$, $q_{1}$, $q_{0}$ and $x_{o}$.

Introducing the notation $\lm_{1} - q_{1}x_{o} \equiv \lm_{2}$ and taking into account 
the definition of the Green function $G_{c}$, eq.(\ref{a17}),
the equations (\ref{a21}) and (\ref{a23}) can be represented as follows:

\bq
\lb{a24}
\[G_{c}(\lm_{2})\] - \[G_{c}(\lm_{1})\] - \fr{1}{2u} (\lm_{1} - \lm_{2}) &=& 0
\\
\nn
\lb{a25}
\int_{\lm_{2}}^{\lm_{1}} d\lm 
\(\[G_{c}(\lm_{2})\] - \[G_{c}(\lm)\] - \fr{1}{2u} (\lm - \lm_{2}) \) &=& 0 \; \; .
\eq
One can easily check that the function 
$\psi(\lm) = \[G_{c}(\lm_{2})\] - \[G_{c}(\lm)\] - \fr{1}{2u} (\lm - \lm_{2})$
has a unique extremum at $\lm = \lm_{o}$ (defined by 
$\[G_{c}^{2}(\lm_{o})\] = \fr{1}{2u}$). Therefore,
the two equations (\ref{a24}) and (\ref{a25}) can be simultaneously satisfied
only when $\lm_{1} = \lm_{2}$, which means that $q_{1}x_{o} = 0$. 
In either of the two cases,
$q_{1} = 0$ or $x_{o} = 0$, we come back to the replica-symmetric structure of the
trial replica matrix $G_{ab}$. 

Quite similar (although more cumbersome) 
calculations show that there are 
also no solutions with more than one step of RSB.
Thus, within the present approximation we cannot have 
solutions with broken replica symmetry in the spin-glass phase.

\sectio{Appendix B}

The saddle-point solutions for non-ferromagnetic states ($m=0$) are defined by 
two order parameters $\lm$ and $\mu$. The stability of these states 
is determined by the signs of the eigenvalues of the corresponding Hessian:

\be
\lb{r45}
\(  
\begin{tabular}{l l}
$\fr{\partial^{2} f}{\partial \mu^{2}} $ & $\fr{\partial^{2} f}{\partial\mu \partial\lm}$ \\
\\
$\fr{\partial^{2} f}{\partial\lm \partial\mu}$ & $\fr{\partial^{2} f}{\partial\lm^{2}} $\\
\end{tabular}
\)
\ee
As usual in the replica theory, one should take into account that in the process
of taking the limit $n\to 0$
the minima of the replica free energy (at  $n > 1$) can turn into
maxima (at $n < 1$). 
The physically relevant states correspond of course to minima of the replica free energy
before taking the limit  $n\to 0$. Around these states all the eigenvalues of the corresponding 
Hessian (composed of the second derivatives of the free energy with respect to {\it all} 
replica components of the order parameters) must be positive. However, if we consider the
expression for the free energy where the limit $n\to 0$ is already taken, the situation
can change. In particular, within the replica-symmetric ansatz for the matrix $\mu_{ab}$, 
eq.(\ref{r25}), the total number of off-diagonal elements (all equal to $-\mu$) is equal to 
$n(n-1)$, and this number becomes {\it negative} for $n < 1$. Therefore, 
the physically relevant state defined by the free energy in eq.(\ref{r32}) (where the limit
$n\to 0$ is already taken) must be the 
{\it maximum} with respect to the parameter $\mu$. 
On the other hand, the total number of diagonal elements
of the matrix $\mu_{ab}$ is equal to $n$, and it remains positive in the limit $n\to 0$. 
Therefore, the physically relevant extremum of the free energy, eq.(\ref{r32}), must be
the {\it minimum} with respect to the parameter $\lm$. 
Thus, in terms of the Hessian, eq.(\ref{r45}), a physically relevant saddle-point
solution must correspond negative (corresponding to the parameter $\mu$) and one
positive (corresponding to the parameter $\lm$) eigenvalue.

By taking the derivatives of the free energy, eq.(\ref{r32}), with respect 
to $\mu$ and $\lm$
and using the saddle-point equations (\ref{r38}) and (\ref{r39})), we find
(for $m=0$)

\bq
\lb{r46}
\fr{\partial^{2} f}{\partial \mu^{2}} &=& \fr{3}{2} g \[G^{2}\]^{2}   
\\
\nn
\lb{r47}
\fr{\partial^{2} f}{\partial\mu \partial\lm} &=& -\2 \[G^{2}\] \(1 + (3g-2u)\[G^{2}\] + 6 g \mu \[G^{3}\] \)       \\
\nn
\lb{r48}
\fr{\partial^{2} f}{\partial\lm^{2}} &=& \2 \[G^{2}\] \(1 + (3g-2u)\[G^{2}\] \) + \mu \[G^{3}\] 
\( 6 g \[G^{2}\] - 1 + 6 g \mu \[G^{3}\] \)    \; \; .      
\eq
For the paramagnetic state ($\mu=0$) we obtain

\bq
\lb{r49}
\fr{\partial^{2} f}{\partial \mu^{2}} &=& \fr{3}{2} g \[G^{2}\]^{2}
\\
\nn
\lb{r50}
\fr{\partial^{2} f}{\partial\mu \partial\lm} &=& -\2 \[G^{2}\] \(1 + (3g-2u)\[G^{2}\] \)
\\
\nn
\lb{r51}
\fr{\partial^{2} f}{\partial\lm^{2}} &=& \2 \[G^{2}\] \(1 + (3g-2u)\[G^{2}\] \) \; \; .
\eq
The two eigenvalues of the Hessian, eq.(\ref{r45}), are then

\be
\lb{r52}
\Lm_{1,2}^{(P)} = \4 \[G^{2}\]\(1 + (6g-2u)\[G^{2}\] \)
          \[1 \pm \sqrt{1 + \fr{4\(1 + (3g-2u)\[G^{2}\] \)}{
          \(1 + (6g-2u)\[G^{2}\]\)^{2}}\(2u\[G^{2}\] - 1\)} \; \] \; \; .
\ee
One of these eigenvalues is  always positive, while the other 
one is negative when $2u\[G^{2}\] > 1$; 
in this case the paramagnetic solution is {\it stable}. 
On the other hand, when $2u\[G^{2}\] < 1$
the second eigenvalue is also positive, 
and the paramagnetic solution is {\it unstable}.
Using eq.(\ref{r34}) we find that the condition 
$2u\[G^{2}\] > 1$ is equivalent to
$\lm(\t) < \lm_{o}$ (eq.(\ref{r42})), which is satisfied when 
$\t > \t_{sg}$. Thus, we can conclude that
the paramagnetic solution is stable only when $\t > \t_{sg}$, 
while for $\t < \t_{sg}$ it becomes {\it unstable}.

We study now the stability of the spin-glass solution. 
Substituting eq.(\ref{r40}) and 
$\[G^{3}\] = C/2\lm_{o}$ into eqs.(\ref{r46})-(\ref{r48}) leads to

\bq
\lb{r53}
\fr{\partial^{2} f}{\partial \mu^{2}} &=& \fr{3g}{8u^{2}}
\\
\nn
\lb{r54}
\fr{\partial^{2} f}{\partial\mu \partial\lm} &=& 
                -\fr{3g}{8u^{2}} \(1 + \fr{2Cu}{\lm_{o}} \mu\)
\\
\nn
\lb{r55}
\fr{\partial^{2} f}{\partial\lm^{2}} &=& 
 \fr{3g}{8u^{2}} \(1 + \fr{2Cu}{\lm_{o}} \mu\)^{2} - \fr{C}{2\lm_{o}} \mu \; \; .
\eq
The two corresponding eigenvalues are then

\be
\lb{r56}
\Lm_{1,2}^{(SG)} = \2 \(f_{\mu\mu}'' + f_{\lm\lm}''\) 
    \[ 1 \pm \sqrt{1 + \fr{3Cg}{4u^{2}\lm_{o}\(f_{\mu\mu}'' 
           + f_{\lm\lm}''\)^{2}} \mu} \; \]   \; \; ,
\ee
where $f_{\lm\lm}''$ and $f_{\lm\lm}''$ are short-hand notations for 
$\partial^{2} f/\partial \mu^{2}$ and $\partial^{2} f/\partial\lm^{2}$,
respectively. From the above expression, one can conclude that 
in the temperature region $\t < \t_{sg}$ where the spin-glass 
solution is physical ($\mu(\t) > 0$, 
eq.(\ref{r43})), there is always one positive and one negative eigenvalue,
 which means that the spin-glass
solution is {\it stable}.

Thus for $\t > \t_{sg}$ the only stable state of the system is paramagnetic, 
while for $\t < \t_{sg}$
the paramagnetic solution becomes unstable and the stable state 
of the system is the spin-glass phase.

\sectio{Appendix C}

For the spin-glass state
in dimensions  $D = 4 - \e$ ($\e \ll 1$)
one again obtains the solutions, eqs(\ref{r40}),(\ref{r41}), where
instead of eqs.(\ref{r33}),(\ref{r34}) one now has

\bq
\lb{r113} 
\[G_{c}\] &=& \int_{|p|<1}\frac{d^{D}p}{(2\pi)^{D}} \fr{1}{p^{2} + \lm} \simeq
             \left. C\(1 + \fr{2}{\e} \lm (1 - \lm^{-\fr{\e}{2}} ) \)\right|_{\lm\ll 1}
             \simeq C
\\
\nn
\lb{r114} 
\[G_{c}^{2}\] &=& \int_{|p|<1}\frac{d^{D}p}{(2\pi)^{D}} \fr{1}{(p^{2} + \lm)^{2}} \simeq
              C\(\fr{2}{\e} (\lm^{-\fr{\e}{2}} - 1) - \lm^{-\fr{\e}{2}}  \) \; \; .
\eq
Then, instead of eq.(\ref{r42}),  eq.(\ref{r40}) leads to

\be
\lb{r115}
\lm_{o} = \[\fr{4Cu (1 - \fr{\e}{2})}{\e + 4Cu}\]^{\fr{2}{\e}} \; \; .
\ee
In the limit $\e \ll u \ll 1$ this result reduces back to eq.(\ref{r42}), 
so that one eventually recovers all the solutions corresponding to
the case $D = 4$ described in section 3.

Let us consider the other limit

\be
\lb{r116}
u \ll \e \ll 1 \; \; .
\ee
In this case eq.(\ref{r115}) yields

\be
\lb{r117}
\lm_{o} \simeq \(\fr{4C}{\e} u\)^{\fr{2}{\e}} \; \; .
\ee
Thus, one can easily see that when $u \ll \e \ll 1$  one
recovers all the solutions described in section 3, in which the value
of $\lm_{o}$ is now given by eq.(\ref{r117}). In particular, it is
this value of $\lm_{o}$ that controls the temperature range of the 
spin-glass phase separating the paramagnetic and the ferromagnetic
phases.

As in section 5.1 where we discussed the validity of the Gaussian
variational approach in dimensions $D = 4$, one should study
in place of of eqs.(\ref{r103}),(\ref{r104}), the following
RG equations:

\bq
\lb{r118}
\fr{d}{d\xi} g(\xi) = \e g - 6 C \(3g - 4u\) g 
                    + O\(g^{3}; g^{2}u; g u^{2}; u^{3}\)
\\
\nn
\lb{r119}
\fr{d}{d\xi} u(\xi) = \e u - 4 C \(3g - 4u\) u 
                    + O\(g^{3}; g^{2}u; g u^{2}; u^{3}\) \; \; .
\eq
These equations show that the higher-order corrections
of the perturbation theory remain irrelevant at spatial scales $R$
bounded by the condition

\be
\lb{r120}
6 C |3g_{0} - 4u_{0}| R^{\e} \ll \e \; \; ,
\ee
which corresponds to the temperature scale

\be
\lb{r121}
\t \gg \t_{*} \sim \(\fr{6 C |3g_{0} - 4u_{0}|}{\e}\)^{\fr{2}{\e}} \; \; .
\ee
Thus, as done in section 5.1, to guarantee that the higher-order  
corrections do not destroy the approximation made in the present 
approach, one needs $\lm_{o} \gg \t_{*}$, or

\be
\lb{r122}
\(\fr{4C}{\e} u\)^{\fr{2}{\e}} \gg 
               \(\fr{6 C |3g_{0} - 4u_{0}|}{\e}\)^{\fr{2}{\e}} \; \; .
\ee
One can easily verify that this condition is satisfied when 

\be
\lb{r123}
\fr{10}{9}  < \fr{g_{0}}{u_{0}} < \fr{14}{9} 
\ee
{\it and} $\e \ll 1$. In other words, we find the 
same restriction on the parameters $g$ and $u$ as in $D=4$, but
in addition one needs the value of $\e = 4 - D$ to be small.


\newpage


\begin{thebibliography}{99}

\bibitem{crit1}  
A.B.Harris, J.Phys. {\bf C 7}, 1671 (1974)\\
A.B.Harris and T.C.Lubensky, Phys.Rev.Lett., {\bf 33}, 1540 (1974)\\
D.E.Khmelnitskii, ZhETF (Soviet Phys. JETP) {\bf 68}, 1960 (1975)\\
G.Grinstein and A.Luther, Phys.Rev. {\bf B 13}, 1329 (1976)\\
R.Folk, Yu.Holovatch and T.Yavors'kii, {\it preprint cond-mat/0106468} (2001) 


\bibitem{numer1} 
W.Selke et al, in: Annual Reviews of Computational Physics {\bf I},
               D.Stauffer, ed. (World Scientific, Singapore, 1994);\\
F.D.A.Reis, S.L.A. de Queiroz, and R.R. dos Santos , Phys.Rev., {\bf B56}, 6013 (1997);\\
H.G.Ballesteros, L.A.Fernandez, V.Martin-Mayor, A.Munoz Sudupe, G.Parisi and J.J.Ruiz-Lorenzo
Phys.Rev., {\bf B58}, 2740 (1998);\\
H.G.Ballesteros, L.A Fernandez, V.Martin-Mayor, A.Munoz Sudupe, G.Parisi and J.J.Ruiz-Lorenzo, 
J.Phys., {\bf A30}, 8379 (1997)


\bibitem{numer2} 
J.Marro, A.Labarta, and J.Tejada, Phys.Rev., {\bf B34}, 347 (1986);\\
D.Chowdhury and D.Stauffer, J.Stat.Phys., {\bf 44}, 203 (1986);\\
P.Braun and M.Fahnle, J.Stat.Phys., {\bf 52}, 775 (1988);\\
T.Holey and M.Fahnle, Phys.Rev., {\bf B41}, 11709 (1990)

\bibitem{Ma-R} S.K.Ma and J.Rudnick, Phys.Rev.Lett., {\bf 40}, 589 (1978)

\bibitem{grif} Vik.S.Dotsenko, J.Phys.A:Math.Gen. {\bf 32}, 2949 (1999);\\
               A.J.Bray, T.McCarthy, M.A.Moore, J.D.Reger and A. P. Young, 
                           Phys.Rev. B {\bf 36}, 2212 (1987);\\
               A.J. McKane, Phys.Rev. B {\bf 49}, 12003 (1994)

\bibitem{rg-rsb}  Vik.S.Dotsenko et al, J.Phys.A: Math.Gen. {\bf 28}, 3093 (1995)\\
                  Vik.S.Dotsenko and D.E.Feldman, J.Phys.A: Math.Gen. {\bf 28}, 5183 (1995)




\bibitem{percol} See for instance, A.L.Korzhenevskii, H-O.Heuer and K.Herrmanns,  
                        J.Phys.A:Math.Gen. {\bf 31}, 927 (1998);\\

\bibitem{DS} D.Sherrington, Phys.Rev., {\bf B22}, 5553 (1980);\\


\bibitem{gva} S.F.Edwards and M.Muthukumar, 
                        J.Chem.Phys., {\bf 89}, 2435 (1988);\\
              S.Korshunov, Phys.Rev., {\bf B48}, 3969 (1993);\\
              T.Giamarchi and P. Le Doussal,
                        Phys.Rev., {\bf B52}, 1242 (1995)
              D.S.Dean and D.Lancaster, Phys.Rev.Lett. {\bf 77}, 3037 (1996) 

\bibitem{gva1} M.Mezard and G.Parisi, 
                        J.Phys.I, {\bf 1}, 809 (1991) 

\bibitem{sg}  M. M\'{e}zard, G. Parisi and M. A. Virasoro,
        "Spin glass theory and beyond" (World Scientific, 1987)

\bibitem{intro} Vik.S.Dotsenko "Introduction to the Replica Theory
                of Disordered Statistical Systems" (Cambridge University
                Press, 2001)

\bibitem{perc} M.Aizenman, Preprint, cond-mat/9609240 
               

\bibitem{rf} 
H.Yoshizawa and D.Belanger, Phys.Rev., B {\bf 30}, 5220 (1984);\\
Y.Shapir, J.Phys. C {\bf 17}, L809 (1984);\\
C.Ro, G.Grest, C.Soukoulist and K.Levin, Phys.Rev., B {\bf 31}, 1682 (1985);\\
J.R.L. de Almeida and R.Bruisma, Phys.Rev., B {\bf 35}, 7267 (1987); \\
M.Guagnelli, E.Marinari and G.Parisi, J.Phys. A, {\bf 26}, 5675 (1993); \\
C. De Dominicis, H.Orland and T.Temesvari, J.Physique I, {\bf 5}, 987 (1996)


\bibitem{bd} E.Br\'ezin and C. De Dominicis, 
        "Interactions of several replicas in the random field Ising model",
        {\it preprint cond-mat/0007457} (2000) 




\end{thebibliography}
\end{document}